\documentclass[10pt, nofootinbib]{revtex4}

\newcommand{\be}{\begin{equation}}
\newcommand{\ee}{\end{equation}}
\newcommand{\bea}{\begin{eqnarray}}
\newcommand{\eea}{\end{eqnarray}}
\newcommand{\bean}{\begin{eqnarray*}}
\newcommand{\eean}{\end{eqnarray*}}

\begin{document}

\title{\Large\bf What is a particle?}

\author{Daniele Colosi${}^a$, Carlo Rovelli${}^b$}

\affiliation{
\small\it ${}^a$Instituto de Matem\'aticas, UNAM,
Campus Morelia, C.P.\,58190, Morelia, Michoac\'an, Mexico\\
\small\it ${}^b$Centre de Physique Th\'eorique de Luminy\footnote{Unit\'e mixte de recherche (UMR 6207) du CNRS et des Universit\'es de Provence (Aix-Marseille I), de la M\'editerran\'ee (Aix-Marseille II) et du Sud (Toulon-Var); laboratoire affili\'e \`a la FRUMAM (FR 2291).}, Case 907, F-13288 Marseille, EU \vspace{0.2cm}
}

\bigskip

\date{\small\today} 

\begin{abstract} 
\noindent 

\noindent Theoretical developments related to the gravitational interaction 
have questioned the
notion of \emph{particle} in quantum field theory (QFT).  For
instance, uniquely-defined particle states do not exist in general, 
in QFT on a curved spacetime. More in general, particle
states are difficult to define in a background-independent quantum 
theory of gravity.  These difficulties
have lead some to suggest that in general QFT should not be
interpreted in terms of particle states, but rather in terms of
eigenstates of {local} operators.  Still, it is not
obvious how to reconcile this view with the empirically-observed
ubiquitous particle-like behavior of quantum fields, apparent
for instance in experimental high-energy physics, or ``particle"-physics.  Here we offer
an element of clarification by observing that already in flat space
there exist --strictly speaking-- {two} distinct notions of particles:
globally defined $n$-particle Fock-states and \emph{local particle
states}.  The last describe the physical objects detected by
finite-size particle detectors and  {are}  eigenstates of local
field operators.  In the limit in which the particle detectors are
appropriately large, global and local
particle states converge in a weak topology (but not in norm).  This
observation has little relevance for flat-space 
theories --it amounts to a reminder that there are boundary
effects in realistic detectors--; but is relevant for gravity. It reconciles the two points of
view mentioned above. More importantly, it provides a definition of
{\em local} particle state that remains well-defined even when the conventional
global particle states are not defined.  This definition plays an
important role in quantum gravity.
\end{abstract}
\maketitle
\section{Introduction}

Is a particle a local or a global object?  On the one hand, a particle
is a local object detected by a local apparatus, such as a
photoelectric detector or a high-energy-experiment bubble chamber.  On
the other hand, the $n$-particle states of quantum field theory (QFT),
namely the eigenstates of the particle-number operator in Fock space,
have a well-known nonlocal character; for instance, they are not
eigenstates of local operators.  There is a tension between these two
facts.

This tension becomes acute when QFT is defined on a
curved spacetime \cite{bd,wald}.  In this contest, the definition of
particle states depends on the choice of a spacelike foliation ``all
over the universe", a choice that has no physical meaning.  In flat 
space, Poincar\'e invariance selects 
preferred foliations and particle states are defined by decomposing
the field into modes and distinguishing positive and negative
frequencies.  On curved spacetime, in general there is no 
symmetry group, no preferred set of 
modes and no preferred decomposition into
positive and negative frequency.  As a consequence, there is no
preferred vacuum state, and the interpretation of the field states in
term of particles appears to be difficult.  

In fact, it is well known that the Poincar\'{e} group plays a central
role in the particle interpretation of the states of the field:
Wigner's celebrated analysis \cite{wigner} has shown that the Fock
particle states are the irreducible representations of the Poincar\'{e} group
in the QFT state space.  The defining properties of the particles,
mass and spin (or helicity), are indeed the invariants of the
Poincar\'e group.  Now, strictly speaking we do not live in a
Poincar\'e invariant region of spacetime: does this means that,
strictly speaking, the world around us has no particles?

Such arbitrariness and ambiguity of the particle concept have led some
theoreticians like Davies to affirms that ``particles do not exist"
\cite{Davies}, a view shared by several relativists.  The idea is that
QFT should be interpreted in terms of local quantities, such as the
integral of energy-momentum-tensor components over finite regions, as
maintained for instance by Wald \cite{wald}.  But this view is not
shared by many other theoreticians, especially (not surprisingly!)  coming
from the ``particle"-physics tradition, who hold that QFT is
fundamentally a formalism for describing processes involving
particles, such as scattering or decays.  A typical example of this
position is Weinberg \cite{weinberg}, who cannot certainly be suspected
of ignoring general relativity.

These difficulties become serious in a background-independent quantum 
context (see for instance \cite{book}). For instance, in loop quantum gravity
\cite{book,lqg}
quantum states of the gravitational field are described in terms of a 
spin network basis. Can we talk about gravitons, or other particle states, 
in loop quantum gravity \cite{gravitons}?  A common view among relativists is that we 
cannot, unless we consider the asymptotically flat context.  But there 
should well be a way of describing what a finite-size 
detector detects, even in a local background-independent theory! 
Indeed, a recent line of development in loop quantum gravity aims at 
computing transition amplitudes between particle states \cite{transitions},
using only {\em finite} spacetime regions, using a formalism 
developed in  \cite{boundary} and in \cite{book}. 
What are those particle states? What is a particle, in a context in which
there is no Poincar\'e invariance and no preferred foliation of a background 
spacetime?

Here we present an observation which may contribute to bring some
clarity, and reconciliate the two points of view.  We address two
related questions: (i) The problem of the local/global nature of
particles.  More precisely: how can an apparatus localized in
spacetime detect a particle state, if a particle state is not an
eigenstate of a local field operator?  And (ii) how can we understand
a local apparatus detecting physical particles in the context of
curved-spacetime QFT, or even in the context of background independent
quantum gravity, where the standard global construction of particle
states is ambiguous?

To address these questions, we observe that if the mathematical
definition of a particle appears somewhat problematic, its operational
definition is clear: particles are the objects revealed by detectors,
tracks in bubble chambers, or discharges of a photomultiplier.  Now,
strictly speaking a particle detector is a measurement apparatus that
cannot detect a $n$-particle Fock state, precisely because it is
localized.  A particle detector measures a local observable field
quantity (for instance the energy of the field, or of a field
component, in some region).  This observables quantity is represented
by an operator that in general has discrete spectrum.  \emph{The
particles observed by the detector are the quanta of this local
operator.} Our key observation is that the eigenstates of this
operator are states of the quantum field that are \emph{similar, but
not identical, to the Fock particle states defined globally}.

Therefore, strictly speaking there are two distinct notions of
particle in QFT. \emph{Local particle states} correspond to the real
objects observed by finite size detectors.  They are eigenstates of
local operators. On the other hand,  \emph{global particle states}, 
such as the Fock particles, namely the eigenstates of the number 
operator in Fock space, can be defined only under certain conditions.  
Global particle states are simpler to define and they approximate well 
the local particle states detected by local measurements.  Therefore 
the global particle states, {\em when they are available}, give a good
approximate description of the physics of the ``real" particles detected
by the detectors. 

In the paper we illustrate the difference between these two classes of
states, and discuss their relation.  The precise sense in which global
states approximate local particle states is subtle.  We show below
that (contrary to what we expected at first) the convergence is not in
the Hilbert space norm, but only in a weaker topology defined by local
observables themselves.

We only deal here with free fields.  This is not necessarily a trivial
context even in flat space, as illustrated for instance by the Unruh
effect \cite{Unruh:db}.  This effect can be understood in terms of a
basis of particle states different from the standard Minkowski one,
indicating that even in flat space there can be ambiguities in the
definition of the notion of particle.  We expect our conclusions to
have general validity also for an interacting theory, but we do not 
venture here into generalizations. 

In flat space, and for inertial observers, the distinction between
global and local particle states is needlessly punctilious, since
physically it boils down to exponentially small correlation effects at
the detector's boundary.  But the distinction is conceptually
important because it indicates that {\em particle states that
describe the physical particles we observe are equally well 
defined in flat space as is curved spacetime, and even in the 
absence of spacetime in a full quantum gravity context}.  
The distinction shows that the global features of the Fock particle 
states have nothing to do with the real observed particles: they 
are an artifact of the simplification taken by approximating a 
truly observed local particle state with easier-to-deal-with 
Fock particles.

This conclusion opens the path
for discussing particle states in a background-independent context.  
Therefore present paper provides the conceptual and technical 
justification for the use of particle states in the research line in the
references \cite{transitions} and, more in general in the boundary 
formulation of quantum field theory \cite{boundary,book}.  The notion of 
particle used in this context is  not 
inconsistent with the standard QFT notion of particle, in spite of 
the local character of the first and the global character of the second.   

In the paper, we first introduce the distinction between local
particles and global particles using a very simple model: two coupled
harmonic oscillators (Section 2).  Then we extend the
construction to field theory in two steps.  First we consider a
sequence of a large number of coupled oscillators (Section 3,4), then we
discuss field theory in Section 5.  In all these cases, we define
global and local particle states and we discuss their relations.  In
Section 6 we summarize our results and we give a general discussion of
the notion of particle in QFT.

\section{Two oscillators}

To begin with, consider two weakly coupled harmonic oscillators $q_{1},
q_{2}$, with unit mass and with the same angular frequency $\omega$;
the dynamics is governed by the hamiltonian
\bea 
H_{0}&=& H_{1}+H_{2} + V = \frac{1}{2}\left(p_1^2 + \omega^2\; {q}_1^2
\right) + \frac{1}{2}\left(p_2^2 + \omega^2\; {q}_2^2 \right)
+\lambda\, q_{1}q_{2}, 
\eea
where $p_{1}, p_{2}$ are the momenta conjugate to $q_{1}, q_{2}$ and,
say, $\lambda\ll\omega^2$.  The state space of the system is ${\cal
H}=L_{2}[R^2,dq_{1},dq_{2}]$ formed by the functions
$\psi(q_{1},q_{2})$.  We can define an orthonormal basis in this
Hilbert space by diagonalizing a complete set of commuting
self-adjoint operators.  Let us choose the set formed by $H_{1}$ and
$H_{2}$.  Call $E_1$ and $E_2$ the eigenvalues of the operators
$H_{1}$ and $H_{2}$ respectively, and $|n_{1},n_{2}\rangle_{\rm loc}$
their common eigenstates.  The reason for the suffix ``loc" will be
clear in a moment.  The integers $n_{1}$ and $n_{2}$ are the quantum
numbers of $E_1$ and $E_2$ and we can interpret them as the number of
quanta in the first and in the second oscillator respectively.  More
precisely, if we measure the energy $H_{1}$ of the first oscillator we
observe that the measurement outcome is quantized:
$E_{1}=\hbar\omega(n_{1}+1/2)$ and $n_{1}$ can be interpreted as the
number of quanta in $q_{1}$.  It is suggestive to call these quanta
``particles".  Call $N_{12}=n_{1}+n_{2}$ the total particle number,
and call $n$-particle states the eigenstates of $H_{1}+H_{2}$. 
Introducing a Fock-like notation, we can write the state with
no-particles also as
\begin{equation} 
|0\rangle_{\rm loc}= |0,0\rangle_{\rm loc};
\label{120}
\end{equation}
the two one-particle states with particles localized on each 
oscillator as
\begin{eqnarray} 
|1\rangle_{\rm loc}= |1,0\rangle_{\rm loc}, 
\label{1loc}\\
|2\rangle_{\rm loc}= |0,1\rangle_{\rm loc}, 
\end{eqnarray}
where the state $|1\rangle_{\rm loc}$ represents a particle on the
first oscillator and the state $|2\rangle_{\rm loc}$ represent a
particle on the second oscillator; and so on.  Notice that, according
to standard Fock-space terminology, any linear combination of
one-particle states
\begin{equation} 
|\psi\rangle_{\rm loc}= c_{1}|1\rangle_{\rm loc}+c_{2}|2\rangle_{\rm loc} 
\label{psiloc}
\end{equation}
is also called a one-particle state. 

Of course the states $|n_{1},n_{2}\rangle_{\rm loc}$ are not
stationary states.  In a perturbation theory in $\lambda$, for
instance, we can compute the probability amplitude for the
particles to ``jump from one oscillator to the other", and so on.  If
we are interested in the stationary states, we need the normal modes of
the system.  These are
\bea 
q_a =\frac{q_{1}+q_{2}}{\sqrt{2}}, \hspace{3em}
q_b = \frac{q_{1}-q_{2} }{ \sqrt{2}}, 
\label{modes}
\eea
with eigenfrequencies
\bea
\omega_{{a}}^2 = \omega^2 + \lambda, \hspace{3em} \omega_{{b}}^2 = 
\omega^2 - \lambda.  
\eea
In terms of these, the hamiltonian factorizes as
\bea
H = H_{a} + H_{b} =
\frac{1}{2}\left(p_a^2 +\omega_{a}q_a^2\right)  +
\frac{1}{2}\left(p_b^2 +\omega_{b}q_b^2\right)  .
\eea
Let $E_{a}$ ($E_{b}$) be the eigenvalues of $H_{a}$ $(H_{b})$, and
denote $|n_{a},n_{b}\rangle_{\rm }$ the common eigenstates of
$H_{a}$ and $H_{b}$.  The number $n_{a}$ ($n_{b}$) is the number of
quanta (or ``particles") in the mode $a$ ($b$).  Call
$N_{ab}=n_{a}+n_{b}$ the total number of these particles in the
system.  For instance the no-particle state is
\begin{equation} 
|0\rangle_{\rm }=  |0,0\rangle_{\rm };
\label{ab0}
\end{equation}
the two one-particle states with particles localized on each
\emph{mode} are
\begin{eqnarray} 
| a\rangle &=& |1,0\rangle, 
\label{a}\\
| b\rangle &=& |0,1\rangle. 
\end{eqnarray}
A generic one-particle state is a state of the form
\begin{equation} 
|\psi\rangle = c_{a} |a\rangle + c_{b} |b\rangle.
\label{psiglob}
\end{equation}
What is the relation between the one-particle states
$|\psi\rangle_{\rm loc}$ defined in (\ref{psiloc}) and the one particle
states $|\psi\rangle$ defined in (\ref{psiglob})?

One may be naively tempted to say that they are the same states,
namely that the two one-particle states $|1\rangle_{\rm loc}$ and
$|2\rangle_{\rm loc}$ (single excitations of the oscillators) are just
linear combinations of the two one-particle states $|a\rangle$ and
$|b\rangle$ (single excitations of the modes).  But this is not the
case.  In the classical theory, $q_{1}$ can be expressed
as the linear combination of the two modes by inverting (\ref{modes}):
\bea 
q_1 =\frac{q_{a}+q_{b}}{\sqrt{2}};
\eea
accordingly, we can choose $c_{a}=c_{b}=1/\sqrt{2}$ in
(\ref{psiglob}), and we obtain a one-particle state which is maximally
concentrated on the first oscillator. Denote it 
\begin{equation} 
|1\rangle= \frac{1}{\sqrt{2}} |a\rangle +
\frac{1}{\sqrt{2}} |b\rangle.
\label{1glo}
\end{equation}
Is this state equal to $|1\rangle_{\rm loc}$?  No, it is not.  If
$\lambda$ is small the two states differ only a little, but they do
differ.  Both states are, in some sense, ``one particle states" and in
both states the ``particle" is on the first oscillator.  However, they
are distinct states.  

We illustrate their difference in two ways.  First, we can simply
write both of them explicitly in the coordinate basis.  It is a simple
exercise to show that
\bea 
\langle q_{1},q_{2}|1\rangle_{\rm loc} &=& 
{\textstyle{\sqrt{\frac{2\omega^3}{\pi}}}}\  q_{1}\ e^{ -
\frac{\omega}{2}(q_1^2+q_2^2) } 
\label{1loc11}
\eea
while
\bea
\langle q_{1},q_{2}|1\rangle_{\rm } &=& 
\textstyle{\frac{(\omega_a \omega_b)^{1/4}}{\sqrt{2\pi}} }
\ \left( \frac{\sqrt{\omega_a}+\sqrt{\omega_b}}{2}
q_{1}+ \frac{\sqrt{\omega_a}-\sqrt{\omega_b}}{2}q_{2}
\right)
\ e^{-\frac{1}{2}\left(\frac{\omega_{a}+\omega_{b}}{2}(q_{1}^2+q_{2}^2)
+\frac{\omega_{a}-\omega_{b}}{2} q_{1}q_{2}\right)}.
\label{1glo11}
\eea
If $\lambda$ is small, $\omega_{a}\sim\omega_{b}\sim\omega$ and the
two states are similar.  In fact, it is easy to compute that their
scalar product is
\bea
{}_{\rm loc}\langle 1|1\rangle &=& 1 
- O(\lambda^2). 
\label{ss}
\eea
which means that, in a sense, the two states are indistinguishable
even at first order in $\lambda$.  Second, we can compare them using
perturbation theory in $\lambda$.  This is instructive because we will
be able to do the same in the context of field theory.  Let us take
$H_{0}=H_{1}+H_{2}$ as unperturbed hamiltonian.  The two states
$|1\rangle_{\rm loc}$ and $|2\rangle_{\rm loc}$ span a degenerate 
eigenspace of $H_{0}$. We must therefore diagonalize $V$ on this 
eigenspace to start perturbation theory. Clearly $V$ is diagonalized 
in this subspace by the two states 
\begin{eqnarray} 
|a\rangle_{0} &=& \frac{|1\rangle_{\rm loc}+|2\rangle_{\rm loc}}{\sqrt 
2} , \\
|b\rangle_{0} &=& \frac{|1\rangle_{\rm loc}-|2\rangle_{\rm loc}}{\sqrt 
2}.  
\end{eqnarray}
We can compute the first order correction to these states using first 
order perturbation theory. It is convenient to use creation and 
annihilation operators 
\begin{eqnarray} 
q_{1,2} &=& \frac{1}{\sqrt{2\omega}}
(a_{1,2}+ a^\dagger_{1,2}),
\\
p_{1,2} &=& \frac{-i\sqrt{\omega}}{\sqrt{2}}
(a_{1,2}-a^\dagger_{1,2})
\end{eqnarray}
in terms of which the perturbation reads
\begin{equation}
V = \frac{\lambda}{2\omega} (a^\dagger_{1}a^\dagger_{2}+a_{1}a_{2}
+a^\dagger_{1}a_{2}+a_{1}a^\dagger_{2}). 
\end{equation}
Notice that the term $a^\dagger_{1}a^\dagger_{2}$ brings out from the
one particle sector, giving the non-vanishing matrix elements
\begin{eqnarray}
\langle 2,1 |V| a \rangle = \hskip4pt
\langle 1,2 |V| a \rangle &=& \frac{\lambda}{2 \omega}\\
\langle 2,1 |V| b \rangle = 
- \langle 1,2 |V| b \rangle &=& \frac{\lambda}{2 \omega}
\end{eqnarray}
To first order in $\lambda$, the hamiltonian eigenstates $| a
\rangle$ and $| b \rangle$ are therefore
\begin{equation} 
|a\rangle\ =\ |a\rangle_{0}\ +\ \frac{\langle 2,1 |V| a
\rangle}{E_{a}-E_{2,1}}\ |2,1\rangle \ +\ \frac{\langle 1,2 |V| a
\rangle}{E_{a}-E_{2,1}}\ |1,2\rangle\ =\  |a\rangle_{0}\ 
-\frac{\lambda}{4\omega^2}\
| 2, 1 \rangle \ -\ \frac{\lambda}{4\omega^2}\ | 1, 2 \rangle
\end{equation}
and
\begin{equation} 
|b\rangle\ =\ |b\rangle_{0}\ +\ \frac{\langle 2,1 |V| b
\rangle}{E_{a}-E_{2,1}}\ |2,1\rangle\ -\ \frac{\langle 1,2 |V| b
\rangle}{E_{a}-E_{2,1}}\ |1,2\rangle\ =\ |b\rangle_{0}\ 
-\frac{\lambda}{4\omega^2}\
| 2, 1 \rangle \ +\ \frac{\lambda}{4\omega^2}\ | 1, 2 \rangle
\end{equation}
And therefore, to first order in $\lambda$
\be 
|1\rangle\ =\ |1\rangle_{\rm loc}\ -\
\frac{\lambda}{\sqrt{8}\, \omega^2}\ |2,1\rangle.
\ee

Thus, the two states (\ref{1loc}) and (\ref{1glo}) are both ``one-particle
states" in which the ``particle" is concentrated on the oscillator
$q_{1}$, but they are distinct states.  They represent two distinct
kinds of one-quantum states, or two distinct kinds of quanta.  We call
$|1\rangle_{\rm loc}$ a \emph{local} particle state, and
$|1\rangle_{\rm }$ a \emph{global} particle state.  They represent the
simplest example of the distinction between these two classes of
states.

More in general, we call ``global particle states" the
eigenstates of the ``global" number operator
\bea 
N_{ab}\ |n_{a},n_{b}\rangle_{\rm } &=& (n_{a}+n_{b})\
|n_{a},n_{b}\rangle\ \ , 
\eea
and we call ``local particle states" the eigenstates of the ``local"
number operator
\bea 
N_{1}\ |n_{1},n_{2}\rangle_{\rm loc} &=& n_{1}\
|n_{1},n_{2}\rangle_{\rm loc} \ \ .
\eea

Let us illustrate the different properties that these states have. 
The state $|1\rangle_{\rm loc}$ is an eigenstate of $H_{1}$, which is
an observable that depends just on $q_{1}$ and its momentum, namely
just on the variable associated to the first oscillator.  If we want
to measure how many local particles are in the first oscillator,
namely to measure $n_{1}$, we can make a measurement that involves
solely variables of the $q_{1}$ oscillator.  In this sense
$|1\rangle_{\rm loc}$ is ``local".

The state $|1\rangle_{\rm }$, on the other hand, describes a single
particle ``on the first oscillator", but is not an eigenstate of
observables that depend on variables of the sole first oscillator. 
This can be seen from the fact that it is a state in which the two
oscillators are (weakly) correlated. The source of these
correlations can be traced to the vacuum state: local particles are
excitation over the local vacuum (\ref{120}) which has no
correlations:
\bea 
\langle q_{1},q_{2}|0\rangle_{\rm loc} &=& 
{\textstyle{\sqrt{\frac{\omega}{\pi}}}}
\ e^{ -\frac{\omega}{2}q_1^2 } 
\ e^{ -\frac{\omega}{2}q_2^2 } = \psi_{0}(q_{1})\,  \psi_{0}(q_{2})
\label{0loc}
\eea
while global  particles are excitations over the 
global vacuum (\ref{ab0})
\bea
\langle q_{1},q_{2}|0\rangle_{\rm } &=& 
\textstyle{\frac{(\omega_a \omega_b)^{1/4}}{\sqrt{\pi}} }
\ e^{-\frac{1}{2}\frac{\omega_{a}+\omega_{b}}{2}q_{1}^2}
\ e^{-\frac{1}{2}\frac{\omega_{a}+\omega_{b}}{2}q_{2}^2}
\ e^{-\frac{1}{2}
\frac{\omega_{a}-\omega_{b}}{2} q_{1}q_{2}},
\label{0glo}
\eea
which does not factorize, and therefore represents vacuum correlations
between the two oscillators.  In Appendix A we give a more precise and
quantitative expression of this correlation.

Notice that $|1\rangle_{\rm loc}$ is not an energy eigenstate, because
of the interaction term $V$, but $|1\rangle_{\rm }$ isn't an energy
eigenstate either, because $ |1,0\rangle_{\rm }$ and $|0,1\rangle_{\rm
}$ have different energies.  Its defining property is just the fact of
being a linear combination of one-quantum excitations of the normal
modes of the system.  What is then the physical relevance of the state
$|1\rangle_{\rm }$?  It is the following: the one-particle Fock states
of QFT are precisely states of the same kind as $|1\rangle_{\rm }$. 
To see this, consider a Fock particle localized in a region $R$.  This
state can be described by means of a function $f(x)$ with compact
support in $R$, as
\bea
|f\rangle_{\rm } &=& \int dk \ \tilde f(k)\ |k\rangle. 
\label{fock1}
\eea
where $\tilde f(k)$ is the Fourier transform of $f(x)$ and the states
$|k\rangle$ are the one-particle Fock states with momentum $k$.  They
are energy eigenstates (with different energies) and they are
single-particle excitations of the normal modes of the system. 
Therefore they are analog to the states $|1,0\rangle_{\rm }$ and
$|0,1\rangle_{\rm }$ of the two-oscillators model.  The linear
combination (\ref{fock1}) is the analog of the linear combination
(\ref{1glo}), which picks the one-particle global state maximally
concentrated in the region chosen (the oscillator $q_{1}$ in the
model, the region $R$ in the QFT).  Thus, Fock particles are global
particles.  No measurement in a finite region $R$ can count those
particles, because Fock particles are not eigenstates of local field
operators, precisely in the same sense in which $|1\rangle_{\rm }$ is
not an eigenstate of an observable localized on the $q_{1}$
oscillator.  If we make a measurement with an apparatus located in the
region $R$, we can count the number of particles the apparatus detect. 
However, these particles are not global particles.  They are local
particles, that can be described by appropriate QFT states which are
close, but not identical, to $n$-particle Fock states, like
$|1\rangle_{\rm loc}$ is close, but not identical to $|1\rangle_{\rm
}$.  Later on, we discuss local particle states, analog to the
$|n_{1},n_{2}\rangle_{\rm loc}$ states, in the context of QFT.

Suppose now the state of the system is $|0\rangle_{\rm }$ and we
measure whether a particle is on the first oscillator by measuring the
energy $E_{1}$.  The probability of not seeing any particle is not
determined by the sole scalar product (\ref{ss}), because we are in
fact tracing over $n_{2}$.  Rather, it is given by
\be 
{\cal P} \ =\  \left| \sum_{n_{2}}\ {}_{\rm loc}\langle 0, n_{2}
|0\rangle_{\rm } \right|^2 
\ =\ 
{}_{\rm }\langle 0|P_{0_{\rm loc}}|0\rangle_{\rm }
\ee
where 
\bea 
P_{0_{\rm loc}}=\sum_{n_2}\ |0,n_{2}\rangle_{\rm loc}\ {}_{\rm loc}\langle
0,n_{2}|
\eea
is the projection on the lowest eigenspace of $H_{1}$.
A straightforward calculation gives 
\be
{\cal P} \ =\ 
{}_{\rm }\langle 0|P_{0_{\rm loc}}|0\rangle_{\rm }
\ =\  1 
-\frac{1}{16}\frac{\lambda^2}{\omega^4} +
O(\lambda^4). 
\label{0p}
\ee
This expression gives a quantitative evaluation of the ``error" that
we make in confusing local particles with global particles: if the
system is in the global vacuum state, there is a probability $1-{\cal
P}$ that a particle detector localized on the first oscillator 
detects a particle.

\section{Chain of oscillators}

As an intermediate step before going to field theory, let us consider
a chain of coupled harmonic oscillators.  This system allows us to
emphasize several important points regarding the relation between
local and global particle states.

We study a system of $n$ harmonic oscillators $\mathbf q=(q^i), 
i=1,\ldots, n$
with the same frequencies $\omega=1$ and coupled by a constant
$\lambda$.  Each oscillator is coupled with its two neighboring (except
the first and the last oscillator that have only one coupling)
\bea 
H&=&  \frac{1}{2}\big(\; |\mathbf p|^2 +|\mathbf q|^2\, 
\big) + \lambda \sum_{i=1}^{n-1}  q^{i}q^{i+1}
\label{H}
\eea
where $|\mathbf q|^2=\sum_i (q^i)^2$.  Notice that we are not
considering a ring but an open chain of oscillators.  Diagonalizing
the hamiltonian of the system we obtain the normal frequencies
\bea
\omega_a= \sqrt{1+2 \lambda \cos \theta_a}, \  \ \; \; \hbox{where} 
\  \ \theta_a = \frac{a \pi}{n+1},\; \; \  \ \hbox{and} \   \ 
a=1,...,n.
\eea
The normal modes $\mathbf Q=(Q_a), a=1,\ldots, n\ $ are given by
$\textbf{Q}= U^{\scriptscriptstyle{(n)}} \textbf{q}$, where $ U^{
\scriptscriptstyle{(n)}}$ is the orthogonal $n \times n$ matrix
\bea
U^{\scriptscriptstyle{(n)}}_{ai}= \sqrt{\frac{2}{n+1}} \sin \left( 
\frac{ai
\pi}{n+1} \right).
\label{matrix}
\eea
The vacuum state is
\bea \langle \mathbf q | 0 \rangle_{\rm } = \prod^{n}_{a=1}
\left(\frac{\omega_a}{\pi}\right)^{1/4} e^{-\frac{1}{2} 
q^i D^{\scriptscriptstyle{(n)}}_{ij} q^j}. 
\label{Pserie}
\eea
where
\bea
D^{\scriptscriptstyle{(n)}}_{ij}= \sum_{a}  U^{\scriptscriptstyle{(n)}}_{ai}
\omega_a U^{\scriptscriptstyle{(n)}}_{aj}
\label{D}
\eea
A basis that diagonalizes $H$ is given by the states $|\mathbf
n\rangle= |n_{1}, \ldots n_n\rangle$ with $n_{a}$
quanta in the $a$-th normal mode.  The number operator is
\bea 
N \  | \mathbf n \rangle_{\rm } = 
\left(\sum_{a=1}^n\ n_{a}\right)\ \  |\mathbf n\rangle
\label{numberop}. 
\eea
Denote $|a\rangle$ the one particle state $|0, \ldots, 1,\ldots,
0\rangle$ in which all normal modes are in the vacuum state except for
the $a$-th mode which is in its first excitation.  The state
\bea 
|i\rangle  = \sum_{a=1}^n\  U^{-1}_{ia}\ |a\rangle
\label{gparticle}
\eea
is the one particle state maximally concentrated on the $i$-th
oscillator.  It is the analog of the global one-particle states
(\ref{1glo}) and (\ref{fock1}).  This is the global one-particle
state, with the particle on the $i$-th oscillator.

Now, consider a partition of the chain in two regions $R_{1}$ and
$R_{2}$.  Let the region $R_{1}$ be formed by the first $n_{1}$
oscillators, and the region $R_{2}$ be formed by the remaining $n_{2}$
oscillators, with $n_{1}+n_{2}=n$.  We write $\mathbf q=(\mathbf
q_{1},\mathbf{q}_{2})$, where $\mathbf q_{1}$ (respectively $\mathbf
q_{2}$ ) is a vector with $n_{1}$ ($n_{2}$) components.  We regard the
first region of oscillators as a generalization of the oscillator
$q_{1}$ in the previous section, and the second region as the analog of
the oscillator $q_{2}$.  The total Hilbert space of the system
factorizes as ${\cal H}={\cal H}_{1}\otimes{\cal H}_{2}$.  We can
rewrite the hamiltonian (\ref{H}) in the form
\bea 
H&=& H_{1}+H_{2}+V 
\\ \nonumber &=& 
\left(\frac{1}{2}\left(|\mathbf p_1|^2 + |\mathbf q_1|^2
\right) + \sum_{i=1}^{n_{1}-1} \lambda \ q_{1}^iq_1^{i+1}\right)
+ \left(\frac{1}{2}\left(|\mathbf p_2|^2 + |\mathbf q_2|^2
\right) + \sum_{i=1}^{n_{2}-1} \lambda \ q_{2}^iq_2^{i+1}\right)
+ \lambda\ q_1^{n_{1}}q_2^1.
\eea
We ask what is a particle, or a quantum excitation of the system,
localized in the region $R_{1}$.  As before, there are two possible
answers.

First, we can consider the vacuum state $|0 \rangle_{\rm }$ and define
the global one-particle states as a linear combination of single
quantum excitations of the normal modes of the system.  In particular,
the linear combination can be chosen to be concentrated in the first
region.  If $i<n_{1}$ is in the first region, (\ref{gparticle}) is a
state representing a global particle state in $R_{1}$.

If we make a measurement in the region $R_{1}$, however --namely if
we measure a quantity that depends only on the variables $\mathbf
q_{1}$ (and their momenta)-- we do not measure a state like
(\ref{gparticle}), because this state is a state where the two regions
are correlated.  More precisely, this state is not an eigenstate of an
observable localized in $R_{1}$.  

Suppose thus that we have only access to observables that are
functions of the oscillator variables $\mathbf q_{1}$ in the first
region.  For concreteness, suppose we measure the energy $H_{1}$
contained in the first region.  Consider eigenvalues and eigenstates
of $H_{1}$ alone.  These are easy to find, since the calculation is
the same as above, only with $n$ replaced by $n_{1}$.  In particular,
we must diagonalize $H_{1}$ in ${\cal H}_{1}$.  For this, \emph{we
need the normal modes of\ $H_{1}$ alone}.  These normal modes are
given by $\mathbf{Q_1}= U^{\scriptscriptstyle{(n_{1})}}
\mathbf{q_{1}}$, where $ U^{\scriptscriptstyle{(n_{1})}}$ is the
orthogonal $n_{1} \times n_{1}$ matrix (\ref{matrix}).  Let
$|0\rangle_{1}$ be the lowest eigenstate of $H_{1}$ in ${\cal H}_{1}$
and $|\mathbf
n_{1}\rangle_{1}= |n_{1}, \ldots n_{n_1}\rangle_{1}$ with $n_{a}$
quanta in the $a$-th normal mode of $H_{1}$ is in its $n_{a}$-th level.   The 
local number operator 
\bea N_{1} \ |\mathbf n \rangle_{1} = \left(\sum_{a=1}^{n_{1}}\
n_{a}\right)\ \ |\mathbf n\rangle_{1} \label{numberop1}.  \eea
is defined on ${\cal H}_1$ and can be extended to the full $\cal H$
(tensoring with the identity in ${\cal H}_2$).  We call local particle
states the eigenstates of the local number operator $N_{1}$.  In
particular, let for instance $|0\rangle_{2}$ be the lowest eigenstate
of $H_{2}$.  Then the states
\bea 
| 0 \rangle_{\rm loc} &=& | 0 \rangle_{1} \otimes | 0 \rangle_{2}, \\
| i \rangle_{\rm loc} &=& | i\rangle_{1}\otimes | 0\rangle_{2} =   \left(\sum_{a=1}^{n_{1}}\ 
(U^{\scriptscriptstyle{(n_{1})}})^{-1}_{ia}\ |a\rangle_{1}
\right)\otimes | 0\rangle_{2} ,
\label{lparticle}
\eea
where, as before, $ |a\rangle_{1}$ is the state with a single
excitation of the $a$-th normal mode of $H_{1}$, are the local vacuum
and the local one particle state -with the particle on the $i$-th
oscillator, associated to the region $R_{1}$.

The two states $| i \rangle$, defined in (\ref{gparticle}), and $| i
\rangle_{\rm loc}$, defined in (\ref{lparticle}), are both
one-particle states where the particle is concentrated on the first
oscillator.  The first is the analog of the localized Fock particle
states used in QFT, the second is a state that can be detected by a
detector localized in the region $R_{1}$.  Similarly, a detector
localized in $R_{1}$ will certainly detect no particles if the system
is in the state $| 0 \rangle_{\rm loc}$, while in QFT we usually
interpret a state where no particle has been measured by a localized
detector as a global vacuum state analogous to $| 0 \rangle$.

\section{Convergence between local and global states}

What is the error we make in ignoring the difference between local and
global states?  Clearly we should expect that the difference
between $| i \rangle_{\rm loc}$ and $| i\rangle$ becomes negligible
if the region $R_{1}$ is sufficiently large, and if $i$ is
sufficiently distant from the boundary of the region $R_{1}$.  This
fact allows us to ignore the difference, and to describe the outcome
of local detectors in terms of global particles without errors in our
predictions.  Thus we expect that
 \be
| i \rangle_{\rm loc}\hspace{2em} \longrightarrow_{\scriptscriptstyle 
n,n_{1}\to\infty}\hspace{2em}   | i\rangle.
\label{giusta}
 \ee
One might expect then that
\be
\langle i| i \rangle_{\rm loc}\hspace{2em} 
\longrightarrow_{\scriptscriptstyle 
n,n_{1}\to\infty}\hspace{2em}   1;
\label{sbagliata}
 \ee
perhaps surprisingly, however, (\ref{sbagliata}) is wrong, as can be
shown by an explicit calculation.  The physical reason is that the two
states $| i \rangle_{\rm loc}$ and $| i\rangle$ are always physical
distinguishable, irrespectively of the size of the regions.  This is
because the second has correlations across the boundary of the two
regions, which are absent in the first.  Indeed (\ref{giusta}) is
correct, but in a more subtle sense that (\ref{sbagliata}): what
converges is just the expectation value of local measurements.  Let us
illustrate this point in some detail. 

Let us illustrate this independence from the size of the regions by
computing the probability $\cal P$ of finding no particles (namely
$H_{1}$ in its lowest eigenstate), if the system is in the global
vacuum $|0 \rangle$, as we did in the case of the two oscillators. 
Let $P_{0_{\rm loc}}=|0\rangle_{1}\ {}_{1}\langle 0|$ be the projector
on the lowest eigenspace of $H_{1}$.  We have, indicating with $\psi_0({\mathbf{q}})$ the global vacuum state in the coordinate representation,
 \bea
 {\cal P} = \left\langle 0 | P_{0_{\rm loc}}| 0\right\rangle = \int d 
 {\mathbf{q}} \ {\psi}_0^* ({\mathbf{q}}) \left( 
  P_{0_{\rm loc}} {\psi_0} \right)({\mathbf{q}})
 \eea
 where
 \be
  \left( 
  P_{0_{\rm loc}} {\psi_0} \right)({\mathbf{q}})= {\psi}_0^* 
 ({\mathbf{q_{1}}}) \int d\mathbf{q}_{1}' \left(\prod^{n}_{a=1} 
 \left(\frac{\omega_a}{\pi}\right)^{1/4}\right) \left(\prod^{n_{1}}_{a=1} 
 \left(\frac{\tilde{\omega}_a}{\pi}\right)^{1/4}\right) 
 e^{-\frac{1}{2}{\mathbf{q}_{1}'}^{T}
 D^{\scriptscriptstyle{(n_{1})}}{\mathbf{q}_{1}'}} 
 e^{-\frac{1}{2}{{\mathbf{q}}}^{'T} 
 D^{\scriptscriptstyle{(n)}}{{\mathbf{q}}'}} 
 \ee
 where $\tilde\omega_{a}$ are the eigenfrequencies of $H_{1}$, and ${q^i}'=q^i$ for $i=n_1+1,...,n$. 
 Performing the integrations we obtain
 \be
\left\langle 0 | P_{0_{\rm loc}}| 0\right\rangle = 
 \left(\prod^{n}_{a=1} \sqrt{\frac{\omega_a}{\pi}}\prod^{n_{1}}_{a=1} 
 \sqrt{\frac{\tilde{\omega}_a}{\pi}}\right)\left(\det\left(\frac{A}{2 
 \pi}\right)\right)^{-1} \left(\det\left(\frac{B}{2 
 \pi}\right)\right)^{-1/2}
 \ee
 where $A$ is a $n_{1}\times n_{1}$ matrix and $B$ a $n_{2} \times n_{2}$ 
 matrix with elements
 \bea
 A_{ij} &=& D^{\scriptscriptstyle{(n_{1})}}_{ij}+ 
 D^{\scriptscriptstyle{(n)}}_{ij}, \   \ \; \; \; \; \; 
 i,j=1,...,n_{1}; \\
 B_{kl} &=& \frac{1}{4} \left(D^{\scriptscriptstyle{(n)}} A^{-1} 
 D^{\scriptscriptstyle{(n)}}\right)_{kl} + 
 D^{\scriptscriptstyle{(n)}}_{kl}, \   \ \; k,l=n_{1}+1,...,n.
 \eea
 Expanding for small values of the coupling $\lambda$ leads to 
 expression
 \bea
 \prod^{n_{1}}_{a=1} \sqrt{\tilde\omega_a} &\approx& 1 - \lambda^2\ 
 \frac{n_{1}-1}{4} 
 \\
 \left(\det\left(\frac{A}{2 }\right)\right)^{-1} &\approx& 1 + 
 \lambda^2 \left(\frac{n_{1}}{2} - \frac{7}{16} \right) \\
 \left(\det\left(\frac{B}{2}\right)\right)^{-1/2} &\approx& 1 + 
 \lambda^2 \left(\frac{n_{2}}{4} - \frac{1}{8} \right)
 \eea
And we obtain, to order
$O(\lambda^2)$ 
\bea {}_{\rm }\langle 0, P_{0_{\rm loc}} \, 0\rangle_{\rm } &=&
\left[1- \lambda^2 \left(\frac{n_{1}+n-2}{4}\right)\right] \left[1+
\lambda^2 \left(\frac{8n_{1}-7}{16}\right) \right] \left[ 1 +
\lambda^2 \left(\frac{n_{2}}{4} - \frac{1}{8} \right)\right] \\
&=& 1  - \frac{\lambda^2}{16}. 
\eea
Observe that this amplitude does not depend on the number of
oscillators in the two chains, in fact, it is equal to the one
obtained in the case $n_1=1$ and $n=2$ in (\ref{0p}). 

A similar result can be obtained for a one-particle state.  Let
$P_{i_{\rm loc}}= | i \rangle_{1}\ {}_{1}\langle i |$ be the projector
on the local one-particle state $ | i \rangle_{1} $ in ${\cal H}_{1}$. 
The quantity $\langle i| P_{i_{\rm loc}}|i \rangle$ gives the
amplitude of seeing the local particle in $i$ with a detector
localized in $R_{1}$ if the state is the global one-particle state
$|i\rangle$.  Its difference from 1 expresses therefore the error we
make in neglecting the difference between local and global states. 
Assuming that $2<i<n_1-1$, a straightforward calculation yields, to
second order in $\lambda$
\bea
\langle i | P_{i_{\rm loc}}|i
\rangle &=& 1  - \frac{\lambda^2}{16},
\label{1p}
\eea
which, again, does not depend on $n_{1}$ or $n$ either. In Appendix A, 
we illustrate this same point using a different technique. 

Why these results are not in contradiction with the possibility of
using global states to approximate local states with arbitrary
accuracy?  Because to find an observable capable of distinguishing
between the local and the global state we have to go to the boundary
of the regions.  The situation is similar to the well-known case of
total charge in QCD: a state with vanishing total charge is always
orthogonal to a state with nonvanishing total charge.  But if a charge
is sufficiently far, its effect is irrelevant on local observables,
hence states with different total charge can converge in the weak
topology defined by local observables \cite{Streater}.  Thus the
correct convergence that describes the physical relation between
global and local states is not in the Hilbert space norm.  It is in a
weak topology determined by the local observables themselves.  As an
example, consider the two point function 
\bea
W(i,j)=\langle 0 | q_i q_j|0\rangle =\langle i | j \rangle 
\eea
expressing the correlation between two states. If $i$ and $j$ are in 
$R_{1}$ and sufficiently far from the border, then this quantity 
converges rapidly to the same quantity computed with local states 
\bea
W_{\rm loc}(i,j) &=& {}_{\rm loc}\langle i | j \rangle_{\rm loc}.  
\eea
In fact, we have 
\bea W(i,j)
 &=& \int d{\mathbf{q}}
\prod^{N}_{j=1} \left(\frac{\omega_j}{\pi}\right)^{1/2} \; q_i \; q_j
\;e^{-{\mathbf{q}}^{T}D^{\scriptscriptstyle{(n)}}{\mathbf{q}}} =
\frac{1}{2} \left(D^{\scriptscriptstyle{(n)}} \right)^{-1}_{ij}.
\eea
Therefore, clearly
\bea
W(i,j)_{\rm loc} &=& \frac{1}{2} 
\left(D^{\scriptscriptstyle{(n_{1})}} \right)^{-1}_{ij}
\eea
But since for $i,j<n_1$ and for small $\lambda$
\bea \left(D^{\scriptscriptstyle{(n)}} \right)^{-1}_{ij} =
\left(D^{\scriptscriptstyle{(n_{1})}} \right)^{-1}_{ij} \approx \left(
1+ \frac{3}{4} \lambda^2 \right) \delta_{i,j} - \frac{1}{2} \lambda \;
\delta_{i,j\pm 1} + \frac{3}{8} \lambda^2 \; \delta_{i,j\pm 2} +
\ldots \eea
it is clear that the two correlation functions $W(ij)$ and $W(ij)_{\rm
loc}$ are equal to arbitrary high order in $\lambda$ if $i$ and $j$
are sufficiently far from the border.  It is then clear that if the
region $R_{1}$ and we stay sufficiently far from the boundary, local
and global particle states are indistinguishable by measuring
local correlations.

\section{Field theory}

Finally, let us get to field theory.  We consider for simplicity a
free scalar field $\phi(x)$ in two spacetime dimensions, confined in a
finite spacial box of size $L$, with reflecting boundary conditions
$\phi(0)=\phi(L)=0$.  Dynamics is governed by the hamiltonian  
\be
H = \frac{1}{2} \int_{0}^L \left(\pi^2-(\partial\phi)^2+m^2\phi^2 \right).
\ee
where $\pi(x)$ is the momentum conjugate to $\phi$. 
Let $k=1,2,\ldots$ label the (discrete) modes of the system and call
$\omega_{k}$ their energy. These are given by 
\bea 
u_{k}(x)= \frac{1}{\sqrt{L \omega_{k}}} \  
\sin \left( \frac{k \pi x}{L}
\right)
\eea
where 
\bea
\omega_{k}^2 = \frac{k^2 \pi^2}{L^2} + m^2. 
\eea
Then 
\bea 
u_k(x,t) = u_{k}(x)\ e^{i \omega_{k} t}
\eea
is a complex solution of the equation of motion.  We can perform a
standard quantization using the operators $a_{k}$ and $a_{k}^\dagger$,
associated to these modes,
\bea 
a_{k} = \int dx\ u_{k}(x) \ \left( \sqrt{2\omega_{k}}\ \phi(x)+i
\sqrt{\frac{2}{\omega_{k}}}\ \pi(x)\right)
\eea
that give
\bea 
\phi(x)&=&\sum_{k} \sqrt{\frac{1}{2\omega_{k}}}\ 
(a_{k}+a^\dagger_{k})\  
u_{k}(x)\\
\pi(x)&=& i\sum_{k} \sqrt{\frac{\omega_{k}}{2}}\ 
(a_{k}-a^\dagger_{k}) \ 
u_{k}(x)
\eea
in terms of which the hamiltonian operator
reads
\be
H = \sum_{k}\ \omega_{k}\ a^\dagger_{k}\ a_{k}.
\ee
Denote $|k\rangle$ the one-particle Fock states with momentum $k$. 
Global one-particle states are linear combinations of the states
$|k\rangle$ 
\be
|f\rangle = \sum_{k}\ f_{k}\ |k\rangle.
\ee
These are eigenstates of the number operator $N=\sum_{k}\
a^\dagger_{k}\ a_{k}$ associated with $H$.  We can say that the
``position" of the (global) particle is determined by the function
\be
f(x,t)=  \sum_{k}\ f_{k}\ u_k(x,t). 
\ee

Now, consider a particle detector of size $R<L$, located in the
region $\cal R$ defined by $x\in [0,R]$.  Say the detector measures
the energy contained in the region $\cal R$ defined by
\be 
H_{R} = \frac{1}{2} \int_{0}^{R}
\left(\pi^2-(\partial\phi)^2+m^2\phi^2 \right).  
\ee
The quantum operator $H_{R}$ can be written in terms of the 
operators $a_{k}$ and $a^\dagger_{k}$, giving 
\be 
H_{R} = \sum_{k,k'} (A_{kk'} a^\dagger_{k}a^\dagger_{k'}+
B_{kk'} a_{k}a_{k'}+
C_{kk'} a^\dagger_{k}a_{k'})
\ee
where the matrices $A, B$ and $C$ are easily computed from the 
eigen-energies and the overlaps
\be 
U_{kk'} =  \frac{1}{2} \int_{0}^{R} \ u_{k}(x)   \ u_{k'}(x).   
\ee
For instance, we have easily
\be 
A_{kk'} = \frac{k^2k'{}^2+m^2-\sqrt{(k^2+m^2)(k'{}^2+m^2)}}
{4\sqrt{\omega_{k}\omega_{k'}}}\ \  U_{kk'} .   
\ee
This matrix does not vanish, hence the operator $H_{1}$ does not
commute with the number operator.  It contains $
a^\dagger_{k}a^\dagger_{k'}$ terms that take out from the one-particle
subspace, and are analogous to the $ a^\dagger_{1}a^\dagger_{2}$
terms that we have encountered in the $V$ term of the two-oscillator
example.  It follows that one-particle Fock states cannot be
eigenstates of this operator.  Therefore when we make a measurement
with a detector that measures the energy $H_{1}$ contained in a finite
region, we project the state on a subspace of Fock space which is not
an $n$-particle Fock state.  

To find the eigenstates of $H_{R}$ we can simply compute the modes
$u^{R}_{k}(x)$ of the field restricted to the $\cal R$ region, as we
did for the chain of oscillators in the previous section.  These have
support on the region $\cal R$ where they are given by
\bea 
u^{R}_{k}(x)= \frac{1}{\sqrt{R \omega^{R}_{k}}} \  
\sin \left( \frac{k \pi x}{R}
\right)
\eea
where
\bea \omega^{R}_{k}{}^2 = \frac{k^2 \pi^2}{R^2} + m^2.  \eea
It is obvious that the eigenstates of $H_{R}$ still
have a particle like structure, given by the excitations of these
modes.  In particular, we call \emph{local} one-particle state all
single excitations of these modes (eigenstates of $H_{R}$) and their
linear combinations.  More precisely, if $a^{R}_{k}$ and
$a^\dagger{}^{R}_{k}$ are the creation and annihilation operators
for the  $u^{R}_{k}(x)$ modes, defined by 
\bea 
a^R_{k} = \int dx\ u^R_{k}(x) \ \left( \sqrt{2\omega^R_{k}}\ \phi(x)+i
\sqrt{\frac{2}{\omega^R_{k}}}\ \pi(x)\right)
\eea
we have 
\bea 
H_R =  \sum_{k}\ \omega^R_{k}\ a^R{}^\dagger_{k}\ a^R_{k}.
\eea
We define the number operator
\be 
N_{R}= \sum_{k} \ a^\dagger{}^{R}_{k}\  a^{R}_{k}
\ee
and we interpret it as the observable giving the number of particles
detected by a detector confined in the region $R$.  The \emph{local}
particle states are defined as the eigenstates of $N_{R}$. 

It is then clear from the discussion that local and global particle
states are distinct.  The firsts represent the states actually
measured by finite size detectors.  The seconds are the ones we
routinely use in QFT calculations.

\subsection{Convergence between local and global particle states}

As we did for the chain of oscillators, it is not hard to show that
local and global particle states do not converge in norm when $L$ and
$R$ are large.  As before, however, the correlation functions defined
by the local particle states converge to the ones defined by the
global particle states.  We shall now show that this is indeed the
case.

To see this, we show that correlation functions in a box of size $L$
converge to the ones computed on Minkowski space as $L$ becomes large. 
It follows that both the correlation functions of the global and local
particles converge to the same value (the correlation of the free
field on Minkowski) for large $L$ and $R$, hence they converge to each
other.  We work below in the case of a field of mass $m$.  In this
case, large $L$ and $R$ means large with respect to the Compton
wavelength $\lambda_{c}=1/m$ of the particle.  The convergence is
exponential in the ratio $\lambda_{c}/R$.  Notice that this implies
the convergence is extremely good for any macroscopic detector of size
$R$.  The massless case is treated in Appendix C. For completeness, in
Appendix B, we discuss also the case of a lattice field theory, which
bridges between the chain of oscillators considered above and field
theory.

We want to compute the Green function for a scalar field with a mass
$m$ quantized in a one-dimensional box of side $L$.  Let us indicate
with $| 0_L \rangle $ the vacuum state of the field, then the two
points function is defined as
\bea \langle 0| \phi(x,t) \phi(x',t') | 0 \rangle &=&
\sum_{k=-\infty}^{+\infty}\frac{1}{{L \omega_{k}}} \sin \left( \frac{k \pi
x}{L} \right) \sin \left( \frac{k \pi x'}{L} \right) e^{i \omega_{k} (t-t')}.
\eea
Notice the Weyrich's formula
\bea \frac{e^{ik \sqrt{r^2+x^2}}}{\sqrt{r^2+x^2}} = \frac{i}{2}
\int^{+ \infty}_{- \infty} e^{i \tau x} H^{(1)}_{0} \left( r
\sqrt{k^2- \tau^2} \right) \, d \tau \eea
valid for $r$ and $x$ real and $0 \leq \arg \sqrt{k^2- \tau^2} < \pi$
and $ 0 \leq \arg k < \pi$.  $H^{(1)}_{0}$ is the Hankel function of
the first kind with index zero.  Using it, we can write
\bea \frac{\exp \left({i(t-t') \sqrt{m^2+(k \pi
/L)^2}}\right)}{\sqrt{m^2+(k \pi /L)^2}} = \frac{i}{2} \int^{+
\infty}_{- \infty} e^{i \tau k \pi /L} H^{(1)}_{0} \left( m
\sqrt{(t-t')^2- \tau^2} \right) \, d \tau \eea
hence
\be \langle 0 | \phi(x,t) \phi(x',t') | 0 \rangle =
\frac{i}{{L}} \sum_{k=0}^{+\infty} \int d \tau\ \sin\frac{k \pi
x}{L}\ \sin\frac{k \pi x'}{L}\ 
\cos\frac{\tau k \pi}{L}\ H^{(1)}_{0} \left( m
\sqrt{(t-t')^2- \tau^2} \right). \ee
We focus the attention on the summation on $k$
\bea \sum_{k=0}^{+\infty} \sin\frac{k \pi x}{L}\ \sin
\frac{k \pi x'}{L}\ \cos\frac{\tau k \pi}{L}
\ &=& \frac{1}{4} \sum_{k=0}^{+\infty} \left[ \cos\left(\frac{k
\pi}{L}(x-x'+ \tau) \right) +\cos\left(\frac{k \pi}{L}(x-x'- \tau)
\right) + \right.  \nonumber\\
&& \left.  - \cos\left(\frac{k \pi}{L}(x+x'+ \tau) \right)
-\cos\left(\frac{k \pi}{L}(x+x'- \tau) \right)\right] .\eea
Consider the sum of the first cosine
\be \sum_{k=0}^{+\infty} \cos\frac{k \pi(x-x'+ \tau)}{L}
 =\frac{1}{2} \left( \frac{1}{1- \exp
\left(i\frac{\pi}{L}(x-x'+\tau+i \epsilon) \right)} + \frac{1}{1- \exp
\left(-i\frac{\pi}{L}(x-x'+ \tau-i \epsilon) \right)} \right) \ee
where a small imaginary part has been added in the exponential in
order to make the summation convergent. We have 
\bea \sum_{k=0}^{+\infty} \sin \left( \frac{k \pi x}{L} \right) \sin
\left( \frac{k \pi x'}{L} \right) \cos \left(\frac{\tau k \pi}{L}
\right) &=& \frac{1}{8} \left( \frac{1}{1-
e^{i\frac{\pi}{L}(x-x'+\tau+i \epsilon)} } + \frac{1}{1-
e^{-i\frac{\pi}{L}(x-x'+\tau-i \epsilon)} } + \right.  \nonumber\\
&& + \frac{1}{1- e^{i\frac{\pi}{L}(x-x'-\tau+i \epsilon)} } +
\frac{1}{1- e^{-i\frac{\pi}{L}(x-x'-\tau-i \epsilon)} } + \nonumber\\
&& - \frac{1}{1- e^{i\frac{\pi}{L}(x+x'+\tau+i \epsilon)} } -
\frac{1}{1- e^{-i\frac{\pi}{L}(x+x'+\tau-i \epsilon)} } + \nonumber\\
&& - \left.  \frac{1}{1- e^{i\frac{\pi}{L}(x+x'-\tau+i \epsilon)} } -
\frac{1}{1- e^{-i\frac{\pi}{L}(x+x'-\tau-i \epsilon)} }\right). \eea
For simplicity we consider the case $t=t'$.  It is useful to express
the Hankel function with the following integral representation
\bea H^{(1)}_{0} \left(i m |\tau | \right) = \frac{2}{i \pi} K_0 ( m
|\tau |) = \frac{2}{i \pi} \int^{\infty}_{0}\frac{\cos (|\tau |
y)}{\sqrt{m^2+y^2}}\ dy \eea
where $K_0$ is the MacDonald function.  The Green function becomes
\bea \langle 0 | \phi(x,t) \phi(x',t) | 0 \rangle = \frac{1}{4L
\pi} \int^{\infty}_{0} dy \frac{1}{\sqrt{m^2+y^2}} \int^{+ \infty}_{-
\infty} d \tau \left( \frac{\cos (|\tau|y)}{1-
e^{i\frac{\pi}{L}(x-x'+\tau+i \epsilon)}} + similar \; terms \right).
\label{2point}
\eea
The integrals in $\tau$ are calculated going in the complex plane of
$\tau$.  The choice of the closure of the contour depends on the sign
of $\tau$ in the exponential of the denominator.  So, for the first
integral, we close the path on the lower half-plane obtaining the
contour $\cal{C}$, yielding
\bea \int^{+ \infty}_{- \infty} d \tau \frac{\cos (|\tau|y)}{1-
e^{i\frac{\pi}{L}(x-x'+\tau+i \epsilon)}} = \int_{\mathcal{C}} d \tau
\frac{\cos (|\tau|y)}{1- e^{i\frac{\pi}{L}(x-x'+\tau+i \epsilon)}}
\eea
The integrand has an infinite number of poles in $\tau = x'-x + 2nL$,
where $n\in N$.  Applying the theorem of residue we find
\bea \int_{\mathcal{C}} d \tau \frac{\cos (|\tau|y)}{1-
e^{i\frac{\pi}{L}(x-x'+\tau+i \epsilon)}} = 2L \sum_n \cos
\left(|x-x'+2nL|y \right).  \eea
And analogous results are found for the other integrals.  Inserting
these results in (\ref{2point}) we arrive at the final expression for
the Green function
\bea \langle 0_L | \phi(x,t) \phi(x',t) | 0_L \rangle &=& \frac{2}{
\pi}\sum_n \left[ K_0 ( m |x-x'+ 2nL|) - K_0 ( m |x+x'+2nL|)\right].
\label{2massive}
\eea

When the size of the box is much greater than Compton wavelength of
the scalar particle, namely $1/m$ in unit $\hbar=c=1$, we can 
distinguish two cases:

$\bullet\ $ If $0 \ll x,x' \ll L$; in this case, in the limit $x \rightarrow x'$
the main contribution to the correlation function comes from the first
MacDonald function in (\ref{2massive}).  In fact, for $n \neq 0$, we
can expand the MacDonald function for large argument, due to the
condition $mL \gg 1$
\bea 
K_0 ( m |x\pm x'+ 2nL|) \approx \sqrt{\frac{\pi}{2( m |x \pm x'+
2nL|)}}e^{- m |x\pm x'+ 2nL|} 
\eea
so that $K_0$ has an exponential decay with the length scale
proportional to the Compton wavelength of the particle.  Therefore
this is a negligible contribution.  In contrast, for $n=0$, the fact
that the MacDonald function diverges when the argument tends to zero
implies
\bea 
K_0(m |x-x'|) \gg K_0(m |x+x'|) 
\eea
Consequently we can write
\bea \lim_{x \rightarrow x'} \langle 0 | \phi(x,t) \phi(x',t) | 0
\rangle = \frac{2}{ \pi}K_0 ( m |x-x'|) \propto \langle 0_{M} | \phi(x,t)
\phi(x',t) | 0_{M} \rangle \eea
where the state $|0_{M} \rangle$ is the vacuum state for the scalar field
quantized in Minkowski spacetime.

$\bullet\ $ If, on the other hand, $x \sim x' \sim 0$ or $x \sim x' \sim L$; in
this case, both Macdonald functions in (\ref{2massive}) contribute
significantly to the correlation function.  In particular, $K_0(m
|x+x'+2nL|)$ is not negligible when $x,x' \approx 0$ for $n=0$, and
when $x,x' \approx L$ when $n=-1$.  This means that when the two
points considered are near the boundary of the box, the correlation
function feels the present of the box, and differs from the
correlation function defined in the whole Minkowski space.

This result illustrate how, if we stay away from the boundary and if
the region $R$ sufficiently large, correlation functions computed
with local states converge to the ones computed with global states.

\section{Conclusion}

We have argued that the particles detected by real measuring apparatus
are local objects, in the sense that they are best represented by QFT
states that are eigenstates of local operators.  We have defined these
states, and denoted them \emph{local particle states}.

This is not what is usually done in QFT, where, instead, we represent
the particles observed in particle detectors by means of a different
set of states: global particle states such as the $n$-particle
Fock states.  

Global particle states provide a good approximation to local particle
states.  The convergence is not in the Hilbert space norm, but in
a weak topology given by local observables.  The approximation is
exponentially good with the ratio of the particle Compton wavelength
with the size of the detector, and the distinction between global and
local states can therefore be safely neglected in concrete
utilizations of QFT. 

However, the distinction is conceptually important because it bears on
three related issues: (i) whether particles are local or global
objects in conventional QFT; (ii) the extent to which the quantum
field theoretical notion of particle can be extended to general
contexts where gravity cannot be neglected; and furthermore, more in
general, (iii) whether particles can be viewed as
the fundamental reality (the ``ontology") described by QFT. Let us
discuss these three issues separately.

(i) The distinction shows that in the context of conventional QFT the
global properties of the particle states are an artifact of an
approximation taken, not an intrinsic property of physically observed
particles.  We view this as a simple and clear answer to the first
question we have addressed: whether particles are local or global
objects in QFT.\footnote{To avoid misunderstanding, let us emphasize
the fact that we are not referring here to limitations of the theory. 
We have assumed here QFT to be exact.  Given a theoretical description
of the world, such as QFT on Minkowski space, it is important to
distinguish three different levels: (i) the world, which is most
presumably not exactly described by the theory (spacetime is curved,
gravity is quantized\ldots), (ii) the ensemble of the empirical data
to which we have access, with their given accuracy; (iii) the theory. 
We have then two distinct problems.  One is the empirical adequacy of
the theory to the ensemble of data.  To be adequate, a theory does not
need to be an exact and complete description of the world; it is
sufficient that it correctly reproduces observations within the
available accuracy.  For instance, we can describe observed waves on a
lake using a theory of waves on a \emph{flat} water surface, or on a
spherical water surface, or on an ellipsoidal water surface\ldots The
first of these options can be perfectly empirically adequate, even is
the Earth isn't actually flat.  Once the theory is chosen, however,
there is then a second issue: the precise identification between
theoretical quantities and empirical data.  \emph{This} is the issue
we have discussed in this paper.  We have placed ourselves in a
regime, or under the assumption, that flat space QFT is empirically
adequate, but we have reconsidered how it should precisely be
interpreted.  Given a particle observed physically, what is the state
of the theory that best describes it?  Our suggestion is that the
usual answer (a Fock particle state) can be replaced with another one
(a local particle state) which is more coherent with the basic rules
of quantum mechanics (because the result of a local measurement has to
be interpreted as an eigenstate of the corresponding operator) and
bears on the possibility of extending QFT methods to more general
contexts.}

(ii) More importantly, the distinction bears on the general validity
of the notion of particle, and on the possibility of utilizing it in
the context in which gravity cannot be neglected.  In so far as
particles are understood as global objects, tied to global
symmetries of spacetime, their utilization outside flat space is
difficult.  In the context of a curved spacetime and, more in general,
in a background independent context where there is no Poincar\'e
invariant background spacetime, the notion of global particle state is
ambiguous, ill defined, or completely impossible to define.  As
mentioned in the introduction, this has lead several theoreticians to
consider interpretations of QFT where particles play no role.  But if
we can understand particles as eigenstates of local operators, with no
reference to global features, then it is clear that we have an
alternative notion of particle that has all the chances to be well
defined in general.  

On a general curved spacetime, a finitely extended detector that
measures the energy $H_{R}$ contained in a finite region of space
$\cal R$ and in a given reference frame, will detect local states
determined by eigenstates of (the Heisenberg operator) $H_{R}$.  These
will have a particle-like structure.  Indeed, they correspond
precisely to the states that best describe flat space QFT measurements
as well (instead of the Fock $n$-particle states).  Thus,
\emph{global} particle states do not generalize, but \emph{local}
particle states, that truly describe what we measure in a bubble
chamber, do.  

The extension of our results to interacting theories should 
be trivial  when the absence of correlation between 
measurements performed in distant regions is assured by the 
cluster decomposition property \cite{weinberg}, but we expect 
the main point to hold in general. It particular, the results
of this paper strongly support the viability of the idea of using
a notion of particle also in the context of the boundary formulation
of quantum field theory \cite{boundary,book}, which is at the
root of the recent calculations of $n$-point functions in quantum gravity
\cite{transitions}.  This formalism provides the possibility to associate state spaces 
with arbitrary hypersurfaces of spacetime by encoding the information 
on the physical processes taking place within a spacetime region into 
the amplitude associated with states on its boundary hypersurface. 

Thus, our conclusion is that the absence of well-defined global
particle states, Poincar\'e invariance, or a preferred vacuum state,
has no bearing on the possibility of interpreting QFT in terms of
particles.  Putting it vividly (but naively) we could say: local
particle detectors detect particles also on a curved or quantized
spacetime.  This leads us to the third issue, which is more
``philosophical", and on which we offer only a few thoughts, without 
any pretension of rigor or completeness. 

(iii) Can we view QFT, in general, as a theory of particles?  Can we
think that reality is made by elementary objects -the particles- whose
interactions are described by QFT? We think that our results suggest
that the answer is partially a yes and partially a no.  

We have argued that \emph{local} particle states can be defined in
general.  In this sense, we share the point of view that QFT can be
interpreted as a theory of particles quite generally.

On the other hand, however, it is clear from the discussion given that
the particles described by the $n$-particle Fock states are
idealizations that do not correspond to the real objects detected in
the detectors.  Moreover, they have unpalatable global properties. 
Therefore it is very difficult to view \emph{them} as the fundamental
objects described by QFT. In particular, there is no reason for
interpreting the Fock basis as ``more physical" or ``more close to
reality" than any other basis in the state space of QFT. Fock
particles aren't more fundamental objects than eigenstates of any
other operators.  If anything, they are \emph{less} fundamental,
because we never measure the Fock number operator.  Interpreting QFT
as the theory of physical objects described by the $n$-particle Fock
states, with their global features, is not only a stumbling bock
toward potentially useful generalizations of flat space QFT, but it
is also in contradiction with what we have learned about the world
with quantum theory.

Can we base the ontology of QFT on local particles?  Yes, but local
particle states are very different from global particle states. 
Global particle states such as the Fock particle states are defined
once and for all in the theory, while each finite size detector
defines its own bunch of local particle states.  Since in general the
energy operators of different detectors do not commute
($[H_{R},H_{R'}]\ne 0$), there is no unique ``local particle basis" in
the state space of the theory, as there is a unique Fock basis. 
Therefore, we cannot interpret QFT by giving a single list of objects
represented by a unique list of states.  In other words, we are in a
genuine quantum mechanical situation in which distinct particle
numbers are complementary observables.  Different bases that
diagonalize different $H_{R}$ operators have equal footing.  Whether a
particle exists or not depends on what I decide to measure.  In such a
context, there is no reason to select an observable as ``more real"
than the others.

The world is far more subtle than a bunch of particles that interact.

\vskip2cm

\appendix

\section*{Appendix A: Density matrix}

In this appendix, we give a different description of the relation
between local and global states, by using a density matrix technique. 

We can obtain all probabilities for measurements performed in a region
$R_{1}$ in terms of a reduced density matrix which is a function of the
sole degrees of freedom in $R_{1}$.  If the state has correlations
between two regions $R_{1}$ and $R_{2}$, the corresponding reduced
density matrix is not the one of a pure state.

Consider the two oscillators system described in Section 2.  The
density matrix of the global vacuum is $\rho=|0\rangle\ \langle 0|$. 
In coordinate space, it reads
\bea
\rho(q_1,q_2,q_1',q_2') &=& \langle 
q_1,q_2|0\rangle\ \langle 0|q_1',q_2'\rangle) \\
&=& \frac{\sqrt{\omega_a \omega_b}}{{\pi}} \exp \left( 
-\frac{1}{2}\frac{\omega_a + \omega_b}{2} 
(q_1^2+q_2^2+q_1'{}^2+q_2'{}^2) -\frac{\omega_a  - \omega_b}{2} (q_1 q_2 
+q_1' q_2') \right).\nonumber
\eea
Tracing on the $q_{2}$ variable yields the reduced density matrix
\bea
\rho_{red} (q_1,q_1') &=& \int dq_2 \; \rho(q_1,q_2,q_1',q_2) \\
&=&  \sqrt{\frac{2\omega_a \omega_b}{\pi (\omega_a + \omega_b)}} \exp 
\left( -\frac{1}{2}\frac{\omega_a  + \omega_b}{2} (q_1^2+q_1'{}^2) 
-\frac{(\omega_a  - \omega_b)^2}{8 (\omega_a + \omega_b)} (q_1  +q_1' 
)^2 \right).
\eea
This density matrix satisfies the properties $Tr(\rho_{red})=1$ as it 
must be for every density matrix and
\bea
Tr(\rho_{red}^2) =  \frac{2\sqrt{\omega_a \, \omega_b}}{\omega_a + 
\omega_b} \approx 1 -\frac{\lambda^2}{4 \omega^4}
\eea
upon the expanding for small $\lambda$, showing that it is a density
matrix of a mixed state, namely that there are vacuum correlations
between the two oscillators.  

Suppose now that we disregard the second oscillator all together and we
consider the $q_{1}$ system alone, with the hamiltonian $H_{1}$.  If 
we measure the energy $E_{1}$ and find the system in the lowest 
eigenstate, then the system will be described by the 
density matrix 
\bea
\rho(q_1,q_1')=
(\omega/\pi)^{1/2} \exp \left( -\omega (q_1^2+q_1'{}^ 2)/2\right)
\eea
of the pure vacuum state of the single oscillator $q_1$.  The relation
between the two density matrices is, to the first nontrivial order in
$\lambda$
\bea
\rho_{red} (q_1,q_1') \approx \rho (q_1,q_1') \; \left(1 - 
\frac{\lambda^2}{8 \omega^4}\left(1-\frac{\omega}{2}(q_1+q_1')^2 
\right) \right).
\eea
Thus, the expectation value of any observable calculated with this
reduced density matrix differs from the one evaluated with the density
matrix of the pure state by a term proportional to $\lambda^2$.  This
gives a precise general evaluation of the difference between the 
local and global vacuum states. 

Let us then repeat this calculation in the case of the chain of
oscillators.  The density matrix for the chain of $n$ oscillators can
be written as
\bea
\rho(\mathbf q, \mathbf q') = \psi_0(\mathbf q) 
\psi^*_0( \mathbf q').
\eea
To obtain the reduced density matrix on the region $R_{1}$ we must
trace over the variables $\mathbf q_2$, obtaining the reduced
matrix
\bea
{\rho}_{red}(\mathbf q_1,\mathbf q_{1}') = \int 
d\mathbf q_2\;{\psi}_0(\mathbf q_1,\mathbf q_2) 
\;{\psi}^*_0(\mathbf q'_1,\mathbf q_2). 
\eea
Computing the integral and expanding for small values of $\lambda$, 
we find
\bea
{\rho}_{red}(\mathbf q_1,\mathbf q_{1}') &=& \pi^{-n_{1}/2} \left( \det 
D^{\scriptscriptstyle{(n)}}\right)^{1/2}\left(\det 
C(D^{\scriptscriptstyle{(n)}}) \right)^{-1/2} \times \\
&& \times \exp \left( -\frac{1}{2}\mathbf{q_1}^T 
D^{\scriptscriptstyle{(n)}} \mathbf{q_1}-\frac{1}{2}\mathbf{q_1}'{}^T 
D^{\scriptscriptstyle{(n_{1})}} 
\mathbf{q_1}_1'+\frac{\lambda^2}{16}(q_1^{n_1}+((q'_1)^{n_1})^2 \right) \nonumber\\
&=&\pi^{-n_{1}/2} \left( \det D^{\scriptscriptstyle{(n)}} 
\right)^{1/2}\left(\det C( D^{\scriptscriptstyle{(n)}}) 
\right)^{-1/2} \times \nonumber\\
&& \times \exp \left( -\frac{1}{2}\mathbf{q_1}^T
D^{\scriptscriptstyle{(n_{1})}} \mathbf{q_1}-\frac{1}{2}\mathbf{q_1}'{}^T
D^{\scriptscriptstyle{(n_{1})}}\mathbf{q_1}'+\frac{\lambda^2}{8}((q_1^{n_1})^2
+((q_1^{n_1})')^
2+q_{n_1} q_{n_1}') \right) \nonumber
\eea
where $C(D^{\scriptscriptstyle{(n)}})$ is the minor of the matrix 
$D^{\scriptscriptstyle{(n)}}$ with first element 
$(D^{\scriptscriptstyle{(n_{1})}})_{n_{1}+1,n_{1}+1}$. It is not difficult to 
check that 
\bea
Tr(\rho)= 
Tr({\rho}_{red})=1
\eea
and $Tr(\rho^2)=1$ as should be for a pure state.  On the other hand,
${\rho}_{red}$ is a density matrix of a mixed state, indeed
\bea
Tr({\rho}_{red}^2) \approx 1 - \frac{1}{8}\lambda^2.
\eea
Its entropy  is
\bea
S= -Tr \left( \rho_{red} \ln (\rho_{red}) \right)  
\ \approx\ \frac{1}{16} \lambda^2.
\eea
This entropy can be calculated expanding the density matrix in basis
of the eigenfunction of the $N$ coupled oscillators and noting that
the leading term in this expansion is the component on the vacuum
state.  This expression gives a quantitative expression of the
difference between the local and global state.  Notice that it does
not go to zero for large $n$ and $n_{1}$.

\section*{Appendix B: Lattice scalar field}

A free scalar field can be modeled as a collection of coupled harmonic
oscillators located on a lattice of space points $x$.  We consider the
dynamical system defined by the hamiltonian
\bea H= \sum_{i=1}^{N}\frac{1}{2}\dot{q}_i^2 +
\frac{1}{2}\sum_{i=1}^{N-1} \omega^2 \left( q_i+ q_{i+1} \right)^2
\eea
where we have fixed the total number $N$ of oscillators.  The normal
frequencies of this system are
\bea \omega_{n} = \sqrt{2} \; \omega \sin \left(\frac{n \pi}{2(N+1)}
\right).  \eea
Using the same notations as in Section 3, we express the
correlation function as
\bea 
W(i,j) = \left\langle \hat{q}_i\hat{q}_j \right\rangle &=& \frac{1}{2}
\left(D\right)^{-1}_{ij} = {\frac{1}{N+1}} \sum_{k} \frac{\sin \left( i \theta_k \right)\sin
\left(j \theta_k \right)}{\sqrt{2} \; \omega \sin \left({\theta_k
}/{2} \right)}
\label{corrlatt}
\eea
where $\theta_k ={k \pi}/{(N+1)}$.  The fraction in (\ref{corrlatt})
can be reexpressed in the form
\bea \frac{\sin(i\theta_k) \sin(j \theta_k)}{\sin(\theta_k/2)}
=\frac{\cos((i-j)\theta_k) - \cos((i+j) \theta_k)}{\sin(\theta_k/2)}. 
\eea
The following relation holds
\bea \frac{\cos((i-1)\theta_k) - \cos(i \theta_k)}{\sin(\theta_k/2)} =
2 \sin((i-1/2)\theta_k); \eea
using which, we have
\bea \frac{\cos((i-j)\theta_k) - \cos((i+j)
\theta_k)}{\sin(\theta_k/2)} = 2 \sum^{i+j}_{p=i-j+1}
\sin((p-1/2)\theta_k).  \eea
Inserting the last expression into (\ref{corrlatt}) and inverting the
sum over $p$ and $k$, we obtain
\bea 
W(i,j) &=&
{\frac{\sqrt{2}}{\sqrt{\omega}(N+1)}} \sum_{k=1}^{N}
\sum^{i+j}_{p=i-j+1} \sin((p-1/2)\theta_k) \\
&=& {\frac{1}{\sqrt{2\omega}(N+1)}}\sum^{i+j}_{p=i-j+1} \cot
\frac{(2p-1) \pi}{4(N+1)}.  \eea
For $N \gg i,j$ we can expand the cotangent function for small
argument, $\cot x = 1/x +o(x^{-3})$:
\bea
W(i,j) &\approx& {\frac{4
(N+1)}{\pi \sqrt{2\omega}(N+1)}}\sum^{i+j}_{p=i-j+1} \frac{1}{2p-1} 
\nonumber\\
&\approx& \frac{2 }{\pi \sqrt{2\omega}} \left[ \Psi_0(i+j+1/2)
-\Psi_0(i-j+1/2)\right], \eea
where $\Psi_0$ is the digamma function.  We obtain a correlation
function independent from the total number $N$ of oscillators. 

\section*{Appendix C: Massless field}

We extend the analysis of the massive scalar field correlation
functions, given in section 5, to the massless case.  The massless
case is more delicate, because of the infrared divergences, and
because of the conformal invariance of the 2d massless theory.  Here
we give explicitly the correlation functions on a finite box, leaving
a detailed physical discussion for further developments.

A massless scalar field in the region $x\in[0,L]$ can be expanded in
the modes
\bea \phi(x,t)= \sum_{k=-\infty}^{+\infty} \left( a_k u_k(x,t) +a_k^*
u_k^*(x,t) \right) \eea
where the functions $u_k(x,t)$ are solutions of the Klein-Gordon
equation.  We want to quantize the field inside a (one dimensional)
box of length $L$ and consequently we impose the following boundary
conditions on the modes of the field:
\bea u_k(0,t)=u_k(L,t)=0; \eea
therefore these functions result to be
\bea u_k(x,t) = \frac{1}{\sqrt{L E_k}} \sin \left( \frac{k \pi x}{L}
\right) e^{i E_k t}; \qquad \hbox{with} \qquad E_k = \frac{|k| \pi}{L}
\eea
The modes $u_k(x,t)$ form a complete orthonormal basis with respect to
the scalar product
\bea \left( u_k(x,t),u_{k'}(x,t) \right) = i \int_0 ^L \left( u_k(x,t)
\partial_t u^* _{k'}(x,t) - [ \partial_t u_k(x,t)]u^* _{k'}(x,t)
\right) dx.  \eea

The quantization of the scalar field promotes the coefficients $a_k$
and $a_k^+$ to the status of annihilation and creation operators
respectively.  These operators define the vacuum state of the field in
the box, called $| 0 \rangle $.  The two-points function can be
written as
\bea \langle 0 | \phi(x,t) \phi(x',t') | 0 \rangle &=&
\sum_{k=-\infty}^{+\infty}\frac{1}{{L E_k}} \sin \left( \frac{k \pi
x}{L} \right) \sin \left( \frac{k \pi x'}{L} \right) e^{i E_k (t-t')}
\\
&=& \sum_{k=0}^{+\infty}\frac{1}{{L E_k}} \left[ \cos \left( \frac{k
\pi (x-x')}{L} \right) - \cos \left( \frac{k \pi (x+x')}{L} \right)
\right] e^{i E_k (t-t')} \nonumber\\
&=& \frac{1}{2} \sum_{k=0}^{+\infty}\frac{1}{{k \pi}} \left[ e^{
\frac{ik \pi (x-x')}{L} }+e^{ \frac{-ik \pi (x-x')}{L} } - e^{
\frac{ik \pi (x+x')}{L} }- e^{ \frac{-ik \pi (x+x')}{L} } \right]
e^{\frac{ik \pi (t-t')}{L} }. \nonumber \eea
We must compute sums of the kind
\bea \sum_{k=1}^{\infty} \frac{e^{iak}}{k} = -\ln (1-{e^{ia}}). 
\eea
Using this,
\bea \langle 0 | \phi(x,t) \phi(x',t') | 0 \rangle &=& \frac{-1}{2
\pi} \left[ \ln (1-{e^{\frac{i\pi(x-x'+t-t') }{L}}})+\ln
(1-{e^{\frac{i\pi(x'-x+t-t') }{L}}}) \right.  \nonumber\\
&-& \left.  \ln (1-{e^{\frac{i\pi(x+x'+t-t') }{L}}})-\ln
(1-{e^{\frac{i\pi(-x-x'+t-t') }{L}}}) \right].
\eea
To simplify the discussion we focus on the
equal time $(t=t')$ correlation function
\bea \langle 0 | \phi(x,t) \phi(x',t) | 0 \rangle &=& \nonumber \\ 
&&\hspace{-9em} = \ \frac{-1}{2
\pi} \left[ \ln (1-{e^{\frac{i\pi(x-x') }{L}}})+\ln
(1-{e^{-\frac{i\pi(x-x') }{L}}}) - \ln (1-{e^{\frac{i\pi(x+x')
}{L}}})-\ln (1-{e^{-\frac{i\pi(x+x') }{L}}}) \right] \nonumber\\
&&\hspace{-9em} =\ 
\frac{-1}{2 \pi} \left[ \ln \left(2-2 \cos \left( \frac{\pi
(x-x')}{L}\right) \right) - \ln \left(2-2 \cos \left( \frac{\pi
(x+x')}{L}\right)\right) \right].
\label{2massless}
\eea
When $0 \ll x,x' \ll L$ and in the limit $x \rightarrow x'$, the main
contribution comes from the first logarithm in (\ref{2massless}):
\be \langle 0 | \phi(x,t) \phi(x',t) | 0 \rangle \ \ \approx\ \ 
\frac{-1}{2 \pi} \ln \left({{\frac{\pi^2(x-x')^2 }{L^2}}} \right)
=
-\frac{1}{\pi} \ln|x-x'| + \frac{1}{\pi} \ln (L/\pi) 
\label{limit2massless}
\ee
Notice that the dependence on the size of the box $L$ appears only as
an additive constant.  To shed light on the meaning of this constant,
recall that in Minkowski spacetime the massless correlation function
is a divergent quantity that needs to be regularized with the
introduction of an infrared cut-off, say $1/N$
\bea \langle 0 | \phi(x,t) \phi(x',t) | 0 \rangle = \frac{-1}{4 \pi}
\ln \left( \frac{(x-x')^2}{N^2} \right) 
=
-\frac{1}{2\pi} \ln|x-x'| + \frac{1}{2\pi} \ln N .
 \eea
One is then interested in physically observable cut-off independent
quantities.  The cut-off dependence is precisely via an additive
constant, namely the same as the dependence on the box size in
(\ref{limit2massless}); in fact the box provides an infrared
regularization of the correlation function for the massless field. 
Viceversa, an infrared cut-off can be interpreted precisely as the
finite detector size, as in this paper.  This relates the problem of
the relation between (massless) local and global particles to the
usual discussion of the relation between infrared divergences and
(independence from) finite size detector effects.

When the points considered are close to the boundaries of the box,
i.e. $x \sim x' \sim 0$ or $x \sim x' \sim L$, the second logarithm in
(\ref{2massless}) is no more negligible with to the first logarithm,
and therefore the correlation function is sensibly different from the
one defined in Minkowski space.

\end{document}